\documentclass[12pt,a4paper]{article}
\usepackage{amsmath}
\usepackage{bm}
\usepackage{amsfonts}
\usepackage{graphicx}
\usepackage[normal]{caption2}
\usepackage{subfigure}
\usepackage{rotating}
\usepackage{citesort}
\usepackage{lscape}
\usepackage{section}
\begin{document}
\renewcommand\arraystretch{1.1}
\setlength{\abovecaptionskip}{0.1cm}
\setlength{\belowcaptionskip}{0.5cm}
%%%%%%%%%%%%%%%%%
\pagestyle{empty}
\newpage
\pagestyle{plain} \setcounter{page}{1} \setcounter{lofdepth}{2}
\begin{center} {\large\bf Phase-space analysis of fragments formed in heavy-ion collisions}\\
\vspace*{0.4cm}
{\bf Sakshi Gautam}\footnote{Email:~sakshigautm@gmail.com} {and \bf Preeti Bansal}\\
{\it  Department of Physics, Panjab University, Chandigarh -160
014, India.\\}
\end{center}
We study the effect of momentum-dependent interactions and a
broader Gaussian on multifragmentation. We also look into the
details of the fragment structure for a broader Gaussian and
momentum-dependent interactions. We find that nucleons forming the
fragments belong to same region of the phase space.
\newpage
\baselineskip 20pt \normalsize
\section{Introduction}
 \par
  The breaking of colliding nuclei into fragments of different sizes has been studied
  for quite a long time. The detailed experimental and theoretical
  studies revealed that the fragmentation is a complex
  process that depends crucially on the reaction inputs like the
  bombarding energy as well as impact parameter \cite{aich1,kumar1,kumar2,willi}.
  Various experimental studies offer an unique opportunity
  to explore the mechanism behind breaking of nuclei into pieces.
  At the same time, heavy ion reaction can also be used to extract
  information about the nature of the matter. Some processes like kaon production \cite{hart1,fuch} give signal about the softer nature
  of the matter, whereas others give indication that matter could be stiffer in
  nature. It is also well accepted that the static equation of state (EOS) cannot describe
  the heavy-ion reaction adequately. The fate of a reaction depends not only on the density,
  but also on the momentum space. Therefore, the momentum-dependent
  interactions play crucial role in the dynamics of a heavy-ion collision. The momentum dependent interactions
  (MDI)
  are found to affect the collective flow drastically \cite{aich2,aich3,sood1}. Due
  to the reduction in the nucleon-nucleon collisions with MDI, the sub-threshold
  particle production is also reduced \cite{aich2} significantly. Some studies are also
   reported in the literature that focuses on the effect of MDI on
  multifragmentation \cite{peilert}. These studies predicted a significant effect
  of MDI on multifragmentation. These effects were more pronounced at peripheral collisions. Unfortunately, no study has been carried out to look into
  the details of the fragment structure using momentum dependent
  interactions. One is interested to understand whether fragments
  are produced due to coalessence or emerge from the
  particular region of the phase space.
  \par
  In addition, interaction range has also a major role to play in the dynamics of heavy-ion collisions \cite{hart2,gaum1,hart3,klak93,kaon1}.
  It has a pronounced effect on the collective flow and on its disappearance as well as
  on multifragmentation \cite{hart2,gaum1}. For example, in Ref. \cite{gaum1} it has been shown that for a broader Gaussian (larger interaction range),
  the energy of disappearance of flow increases. Similarly, there is a significant effect
  of interaction range on the fragmentation as well. In Ref. \cite{hart2} it has been shown that
  a broader Gaussian leads to reduced fragments. But the details of fragment structure for a broader
  Gaussian was never studied. We, therefore aim to
address\\
  1. The effect of MDI on the fragment structure and\\
  2. The effect of interaction range and to look if fragments then produced belong to certain space or just
   produced in the reaction without pre-selection.\\
  This study is carried out within the framework of quantum
  molecular dynamics (QMD) model
\par
\section{The Formalism}
\subsection{Quantum Molecular dynamics (QMD) model}
\par
We describe the time evolution of a heavy-ion reaction within the
framework of Quantum Molecular Dynamics (QMD) model \cite{aich2}
which is based on a molecular dynamics picture. This model has
been successful in explaining collective flow \cite{sood2},
elliptic flow \cite{kumar3}, multifragmentation \cite{dhawan} as
well as dense and hot matter \cite{fuchs}. Here each nucleon is
represented by a coherent state of the form
\begin{equation}
\phi_{\alpha}(x_1,t)=\left({\frac {2}{L \pi}}\right)^{\frac
{3}{4}} e^{-(x_1-x_{\alpha }(t))^2}
e^{ip_{\alpha}(x_1-x_{\alpha})} e^{-\frac {i p_{\alpha}^2 t}{2m}}.
\label {e1}
\end{equation}
Thus, the wave function has two time dependent parameters
$x_{\alpha}$ and $p_{\alpha}$.  The total n-body wave function is
assumed to be a direct product of coherent states:
\begin{equation}
\phi=\phi_{\alpha}
(x_1,x_{\alpha},p_{\alpha},t)\phi_{\beta}(x_2,x_{\beta},
p_{\beta},t)....,         \label {e2}
\end{equation}
where antisymmetrization is neglected. One should, however, keep
in the mind that the Pauli principle, which is very important at
low incident energies, has been taken into account. The initial
values of the parameters are chosen in a way that the ensemble
($A_T$+$A_P$) nucleons give a proper density distribution as well
as a proper momentum distribution of the projectile and target
nuclei. The time evolution of the system is calculated using the
generalized variational principle. We start out from the action
\begin{equation}
S=\int_{t_1}^{t_2} {\cal {L}} [\phi,\phi^{*}] d\tau, \label {e3}
\end{equation}
with the Lagrange functional
%%%%%%%%%%%%%%%%%%%%%%%%%%%%%%%%%%%%%%%%%%%%%%%%%%%%555
\begin{equation}
{\cal {L}} =\left(\phi\left|i\hbar \frac
{d}{dt}-H\right|\phi\right), \label {e4}
\end{equation}
%%%%%%%%%%%%%%%%%%%%%%%%%%%%%%%%%%%%%%%55
where the total time derivative includes the derivatives with
respect to the parameters. The time evolution is obtained by the
requirement that the action is stationary under the allowed
variation of the wave function
%%%%%%%%%%%%%%%%%%%%%%%%%%%5555
\begin{equation}
\delta S=\delta \int_{t_1}^{t_2} {\cal {L}} [\phi ,\phi^{*}] dt=0.
\label{e5}
\end{equation}
%%%%%%%%%%%%%%%%%%%%%%%%%%%%%%%
If the true solution of the Schr\"odinger equation is contained in
the restricted set of wave function
$\phi_{\alpha}\left({x_{1},x_{\alpha},p_{\alpha}}\right),$ this
variation of the action gives the exact solution of the
Schr\"odinger equation. If the parameter space is too restricted,
we obtain that wave function in the restricted parameter space
which comes close to the solution of the Schr\"odinger equation.
Performing the variation with the test wave function (2), we
obtain for each parameter $\lambda$ an Euler-Lagrange equation;
\begin{equation}
\frac{d}{dt} \frac{\partial {\cal {L}}}{\partial {\dot
{\lambda}}}-\frac{\partial \cal {L}} {\partial \lambda}=0.
\label{e6}
\end{equation}
For each coherent state and a Hamiltonian of the form, \\

$H=\sum_{\alpha}
\left[T_{\alpha}+{\frac{1}{2}}\sum_{\alpha\beta}V_{\alpha\beta}\right]$,
the Lagrangian and the Euler-Lagrange function can be easily
calculated \cite{aich2}
\begin{equation}
{\cal {L}} = \sum_{\alpha}{\dot {\bf x}_{\alpha}} {\bf
p}_{\alpha}-\sum_{\beta} \langle{V_{\alpha
\beta}}\rangle-\frac{3}{2Lm}, \label{e7}
\end{equation}
\begin{equation}
{\dot {\bf x}_{\alpha}}=\frac{{\bf
p}_\alpha}{m}+\nabla_{p_{\alpha}}\sum_{\beta} \langle{V_{\alpha
\beta}}\rangle, \label {e8}
\end{equation}
\begin{equation}
{\dot {\bf p}_{\alpha}}=-\nabla_{{\bf x}_{\alpha}}\sum_{\beta}
\langle{V_{\alpha \beta}}\rangle. \label {e9}
\end{equation}
Thus, the variational approach has reduced the n-body
Schr\"odinger equation to a set of 6n-different equations for the
parameters which can be solved numerically. If one inspects  the
formalism carefully, one finds that the interaction potential
which is actually the Br\"{u}ckner G-matrix can be divided into
two parts: (i) a real part and (ii) an imaginary part. The real
part of the potential acts like a potential whereas imaginary part
is proportional to the cross section.
%%%%%%%%%%%%%%%%%%%%%%%%%%%%%%%%%%%

In the present model, interaction potential comprises of the
following terms:
\begin{equation}
V_{\alpha\beta} = V_{loc}^{2} + V_{loc}^{3} + V_{Coul} + V_{Yuk} +
V_{MDI}, \label {e10}
\end {equation}
$V_{loc}$ is the Skyrme force whereas $V_{Coul}$, $V_{Yuk}$ and
$V_{MDI}$ define, respectively, the Coulomb, Yukawa and momentum
dependent potentials. The Yukawa term separates the surface which
also plays the role in low energy processes like fusion and
cluster radioactivity \cite{puri}. The expectation value of these
potentials is calculated as
%%%%%%%%%%%%%%%%%%%%%%%%%%%%%%%%%%%%55
\begin{eqnarray}
V^2_{loc}& =& \int f_{\alpha} ({\bf p}_{\alpha}, {\bf r}_{\alpha},
t) f_{\beta}({\bf p}_{\beta}, {\bf r}_{\beta}, t)V_I ^{(2)}({\bf
r}_{\alpha}, {\bf r}_{\beta})
\nonumber\\
&  & \times {d^{3} {\bf r}_{\alpha} d^{3} {\bf r}_{\beta}
d^{3}{\bf p}_{\alpha}  d^{3}{\bf p}_{\beta},}
\end{eqnarray}
%%%%%%%%%%%%%%%%%%%%%%%%%%%%%%%%%%%%%%%%5555
\begin{eqnarray}
V^3_{loc}& =& \int  f_{\alpha} ({\bf p}_{\alpha}, {\bf
r}_{\alpha}, t) f_{\beta}({\bf p}_{\beta}, {\bf r}_{\beta},t)
f_{\gamma} ({\bf p}_{\gamma}, {\bf r}_{\gamma}, t)
\nonumber\\
&  & \times  V_I^{(3)} ({\bf r}_{\alpha},{\bf r}_{\beta},{\bf
r}_{\gamma}) d^{3} {\bf r}_{\alpha} d^{3} {\bf r}_{\beta} d^{3}
{\bf r}_{\gamma}
\nonumber\\
&  & \times d^{3} {\bf p}_{\alpha}d^{3} {\bf p}_{\beta} d^{3} {\bf
p}_{\gamma}.
\end{eqnarray}
%%%%%%%%%%%%%%%%%%%%%%%%%%%%%%%%%%%%%%%%%%%%%
where $f_{\alpha}({\bf p}_{\alpha}, {\bf r}_{\alpha}, t)$ is the
Wigner density which corresponds to the wave functions (eq. 2). If
we deal with the local Skyrme force only, we get
%%%%%%%%%%%%%%%%%%%%%%%%%%%%%%%%%%%%%%%55
{\begin{equation} V^{Skyrme} = \sum_{{\alpha}=1}^{A_T+A_P}
\left[\frac {A}{2} \sum_{{\beta}=1} \left(\frac
{\tilde{\rho}_{\alpha \beta}}{\rho_0}\right) + \frac
{B}{C+1}\sum_{{\beta}\ne {\alpha}} \left(\frac {\tilde
{\rho}_{\alpha \beta}} {\rho_0}\right)^C\right].
\end{equation}}
%%%%%%%%%%%%%%%%%%%%%%%%%%%%%%%%%%%%%%%%%%%

 Here A, B and C are the Skyrme parameters which are
defined according to the ground state properties of a nucleus.
Different values of C lead to different equations of state. A
larger value of C (= 380 MeV) is often dubbed as stiff equation of
state.The finite range Yukawa ($V_{Yuk}$) and effective Coulomb
potential ($V_{Coul}$) read as:
\begin{equation}
V_{Yuk} = \sum_{j, i\neq j} t_{3}
\frac{exp\{-|\textbf{r}_{\textbf{i}}-\textbf{r}_{\textbf{j}}|\}/\mu}{|\textbf{r}_{\textbf{i}}-\textbf{r}_{\textbf{j}}|/\mu},
\end{equation}
\begin{equation}
V_{Coul} = \sum_{j, i\neq
j}\frac{Z_{eff}^{2}e^{2}}{|\textbf{r}_{\textbf{i}}-\textbf{r}_{\textbf{j}}|}.
\end{equation}
\par
The Yukawa interaction (with $t_{3}$= -6.66 MeV and $\mu$ = 1.5
fm) is essential for the surface effects. The momentum-dependent
interactions (MDI) are obtained by parameterizing the momentum
dependence of the real part of the optical potential. The final
form of the potential reads as \cite{aich2}
\begin{equation}
U^{MDI}\approx
t_{4}\ln^{2}[t_{5}(\textbf{p}_{1}-\textbf{p}_{2})^{2}+1]\delta(\textbf{r}_{1}-\textbf{r}_{2}).
\end{equation}
Here $t_{4}$=1.57 MeV and $t_{5}$=5$\times$10$^{-4}$ MeV$^{-2}$. A
parameterized form of the local plus MDI potential is given by
\begin{equation}
U=\alpha(\frac{\rho}{\rho_{o}})+\beta(\frac{\rho}{\rho_{o}})+\delta
\ln^{2}[\epsilon(\rho/\rho_{o})^{2/3}+1]\rho/\rho_{o}.
\end{equation}
The parameters $\alpha$, $\beta$, $\gamma$, $\delta$ and
$\epsilon$ are listed in Ref. \cite{aich2}. The momentum-dependent
part of the interaction acts strongly in the cases where the
system is mildly excited. In this case, the MDI is reported to
generate a lot more fragments compared to the static equation of
state. The relativistic effect does not play role in low incident
energy of present interest \cite{lehm}.
\par
The phase space of the nucleons is stored at several time steps
and this is clustered using minimum snapping tree method that
binds the nucleons if they are closer than 4 fm.
\par
\section{Results and Discussion}
 We simulated the reactions of
$^{58}$Ni+$^{58}$Ni and $^{197}$Au+$^{197}$Au for 100 MeV/nucleon
at central ($\hat{\textrm{b}}$=0.2) and peripheral
($\hat{\textrm{b}}$=0.8) colliding geometry. For the present
study, we used stiff (Hard), soft (Soft), soft with momentum
dependent interactions (SMD) equations of state. The standard
energy-dependent Cugnon cross section is used along with two
different Gaussian widths, i.e., L = 1.08 fm$^{2}$
(L$^{\textrm{norm}}$) and 2.16 fm$^{2}$ (L$^{\textrm{broad}}$).
%%%%%%%%%%%%%%%%%%%%%%%%%%%%%%%%%%%%%%%%%%%%
\begin{figure}[!t]
\centering
 \vskip 1cm
\includegraphics[angle=0,width=16cm]{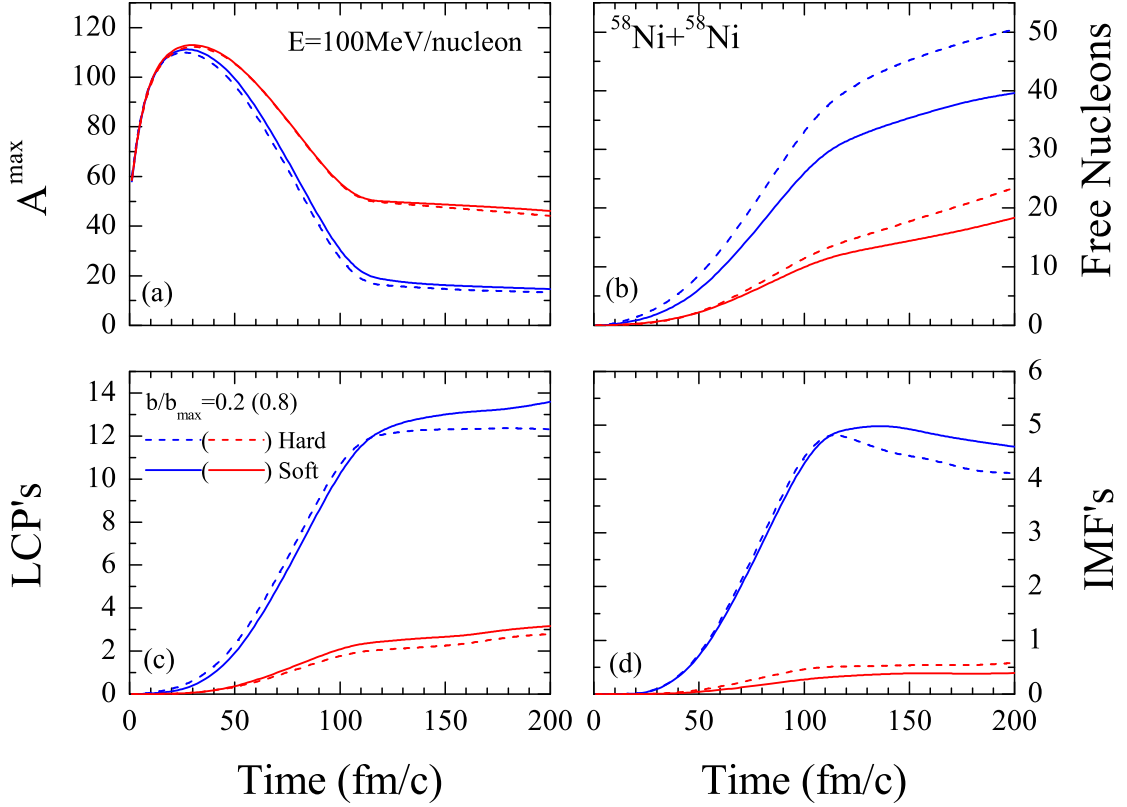}
 \vskip -0cm \caption{ The time evolution of A$^{\textrm{max}}$,
 free nucleons, LCPs and IMFs for the reaction of $^{58}$Ni+$^{58}$Ni at incident energy of
 100 MeV/nucleon at central ($\hat{b}$ = 0.2) and peripheral ($\hat{b}$ = 0.8) for Soft and Hard EOS. Lines are explained
 in the text.}\label{fig1}
\end{figure}
%%%%%%%%%%%%%%%%%%%%%%%%%%%%%%%%%%%%%%%%%%%%%%%%%%

\begin{figure}[!t]
\centering \vskip 1cm
\includegraphics[angle=0,width=16cm]{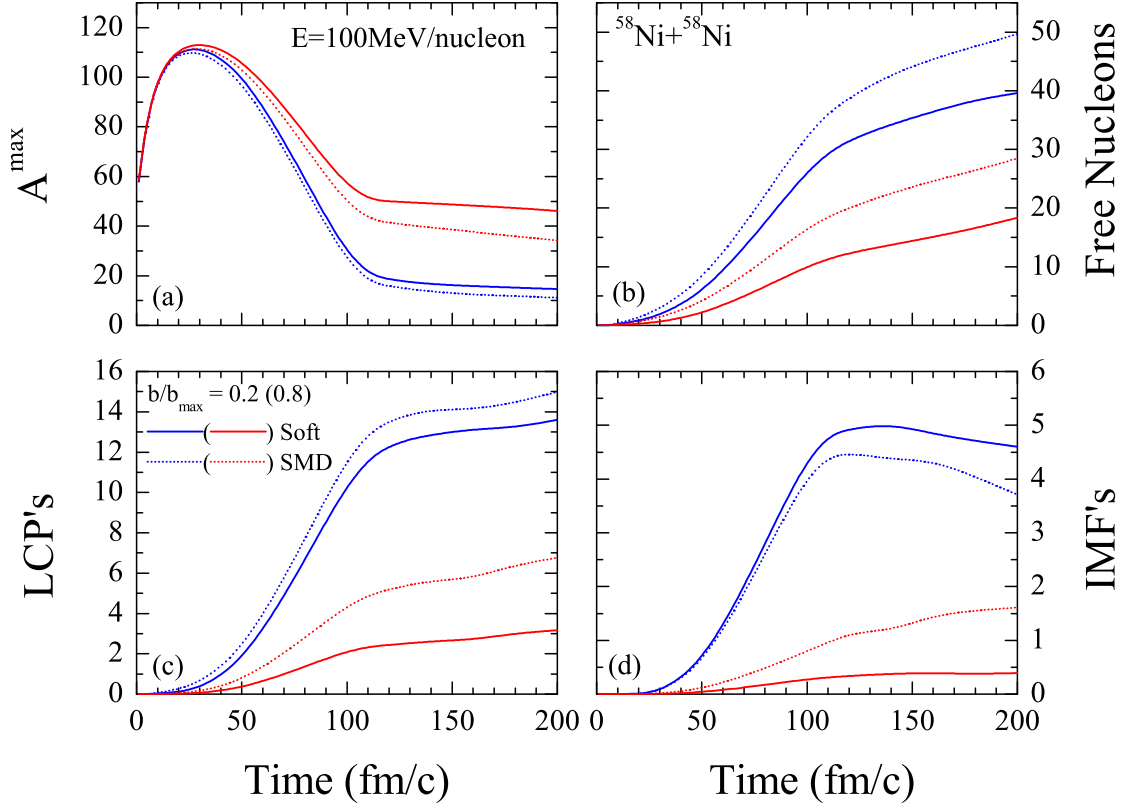}
\vskip -0cm \caption{Same as Fig. 1, but for Soft and
SMD.}\label{fig2}
\end{figure}

\begin{figure}[!t]
\centering \vskip 1cm
\includegraphics[angle=0,width=16cm]{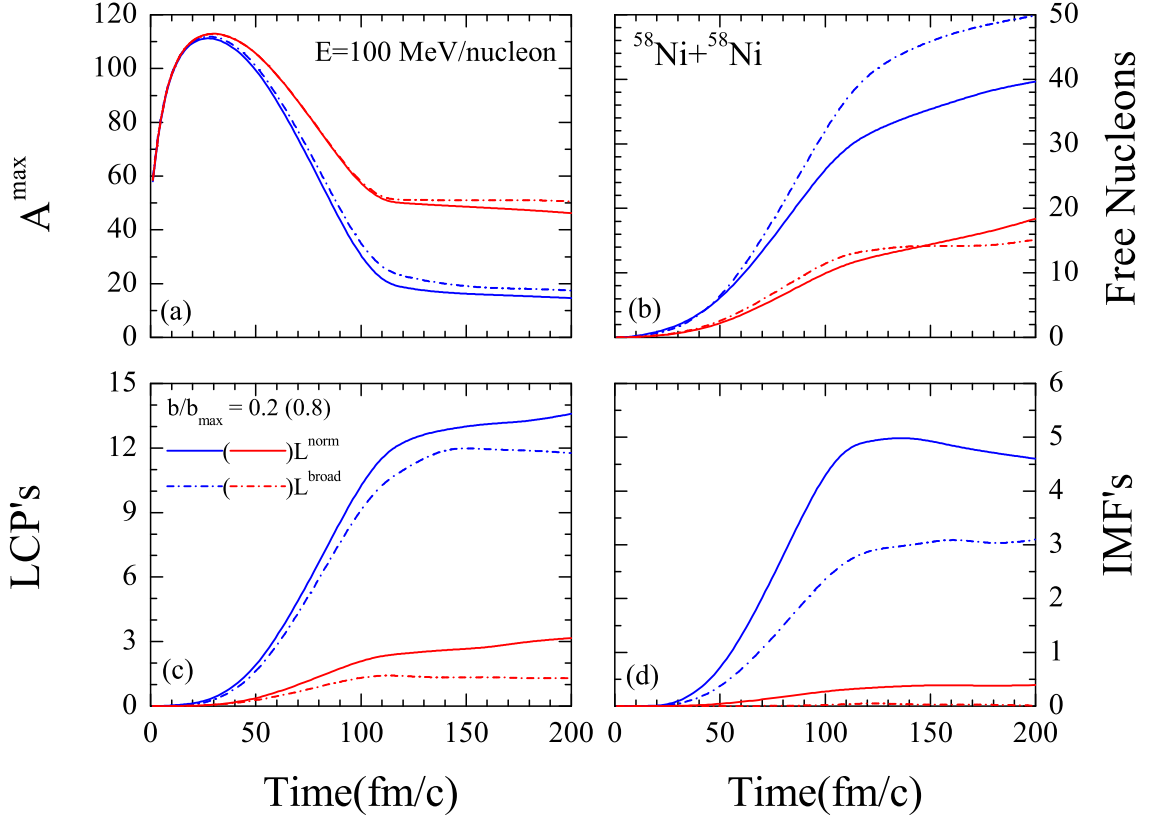}
\vskip -0cm \caption{Same as Fig. 1, but for L$^{norm}$ and
L$^{broad}$.}\label{fig3}
\end{figure}

\par
\begin{figure}[!t]
\centering \vskip 1cm
\includegraphics[angle=0,width=16cm]{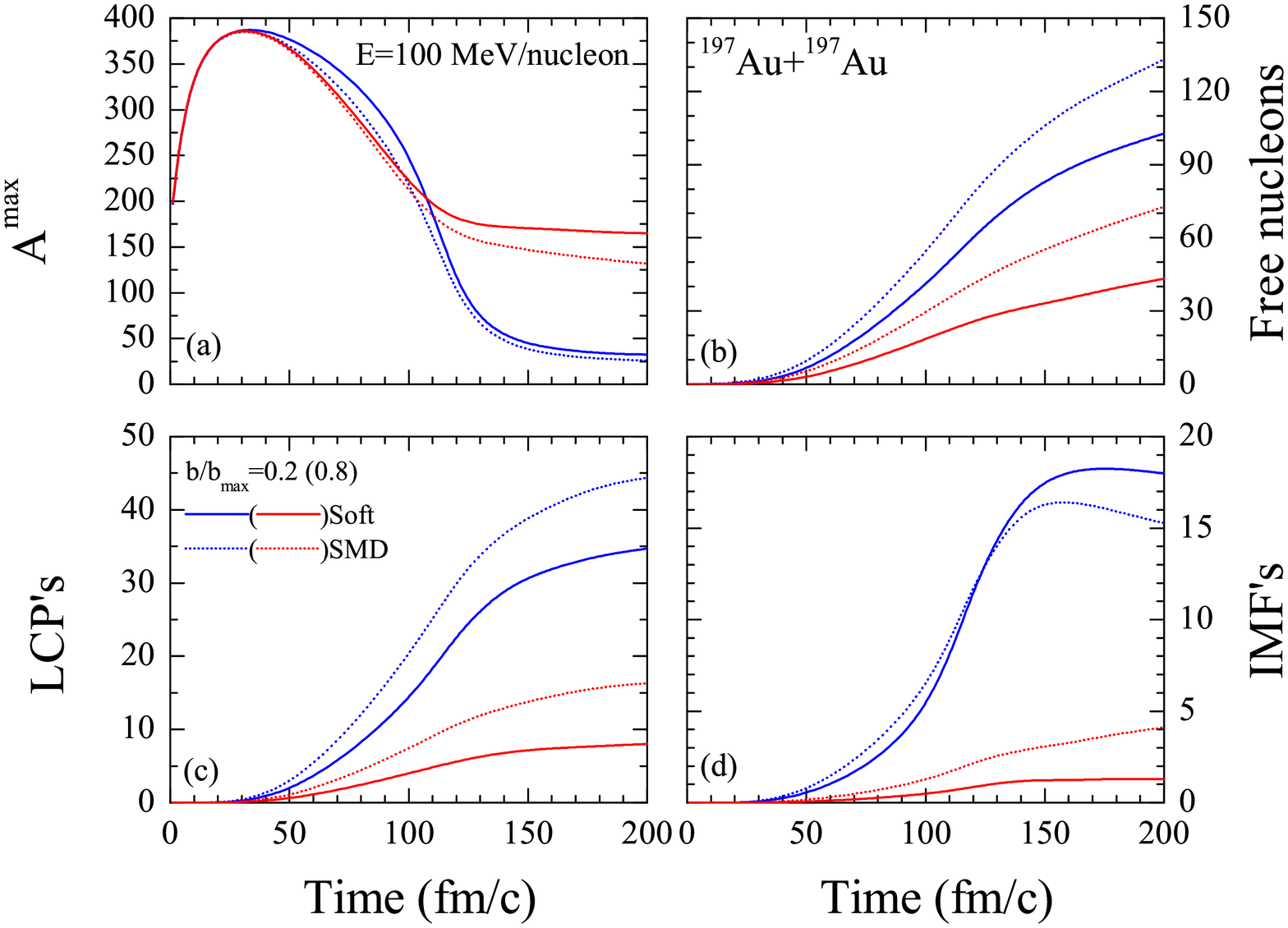}
\vskip -0cm \caption{Same as Fig. 2, but for the reaction of
$^{197}$Au+$^{197}$Au.}\label{fig3}
\end{figure}

\par

In Fig. 1, we display the time evolution of A$^{\textrm{max}}$,
free nucleons, LCPs (2 $\leq$ A $\leq$ 4) and IMFs (5 $\leq$ A
$\leq$ A/3) for the reaction of $^{58}$Ni+$^{58}$Ni at $\hat{b}$
=0.2 and incident energy of 100 MeV/nucleon. The purpose of
showing different mass windows is to identify the different
phenomena that may appear in one window but not in other mass
range. The A$^{\textrm{max}}$ will give a possibility to look for
the fusion (if any), whereas the emission of free nucleons will
show the disassembly and hence vaporization of the nuclear matter.
For the central collision of $\hat{b}$ =0.2 (blue lines)  we see
from Fig.1(a) that A$^{\textrm{max}}$ first increases with time,
reaches maximum (about 116 which is A$^{\textrm{projectile}}$ +
A$^{\textrm{target}}$ ) at about 20-40 fm/c when the matter is
highly compressed and then decreases during the later stages at
about 120 fm/c. The effect of EOS is negligible on
A$^{\textrm{max}}$ (solid and dashed lines) as predicted in Ref.
\cite{kumar1}. From Fig. 1(b), (c), and (d), we find that free
nucleons, LCPs, and IMFs increases with time. This is because the
excited compound nucleus decays by the emission of nucleons and
fragments. As a result, free nucleons, LCPs, and IMFs display a
constant rise in their multiplicities. The constant emission of
free nucleons with time suggests that hot fragments are cooling
down. The emission of free nucleons, LCPs, and IMFs starts at
around 50 fm/c. We also find a significant effect of EOS on the
production of free nucleons, LCPs, and IMFs. We find that number
of LCPs/IMFs is larger in the case of soft EOS compared to hard
EOS (see blue and red lines). This is because of the fact that
soft matter can be easily compressed. As a result, a greater
density can be achieved, which in turn leads to the large number
of IMFs compared to that in hard case.
\par

\begin{figure}[!t]
\centering
 \vskip 1cm
\includegraphics[angle=0,width=16cm]{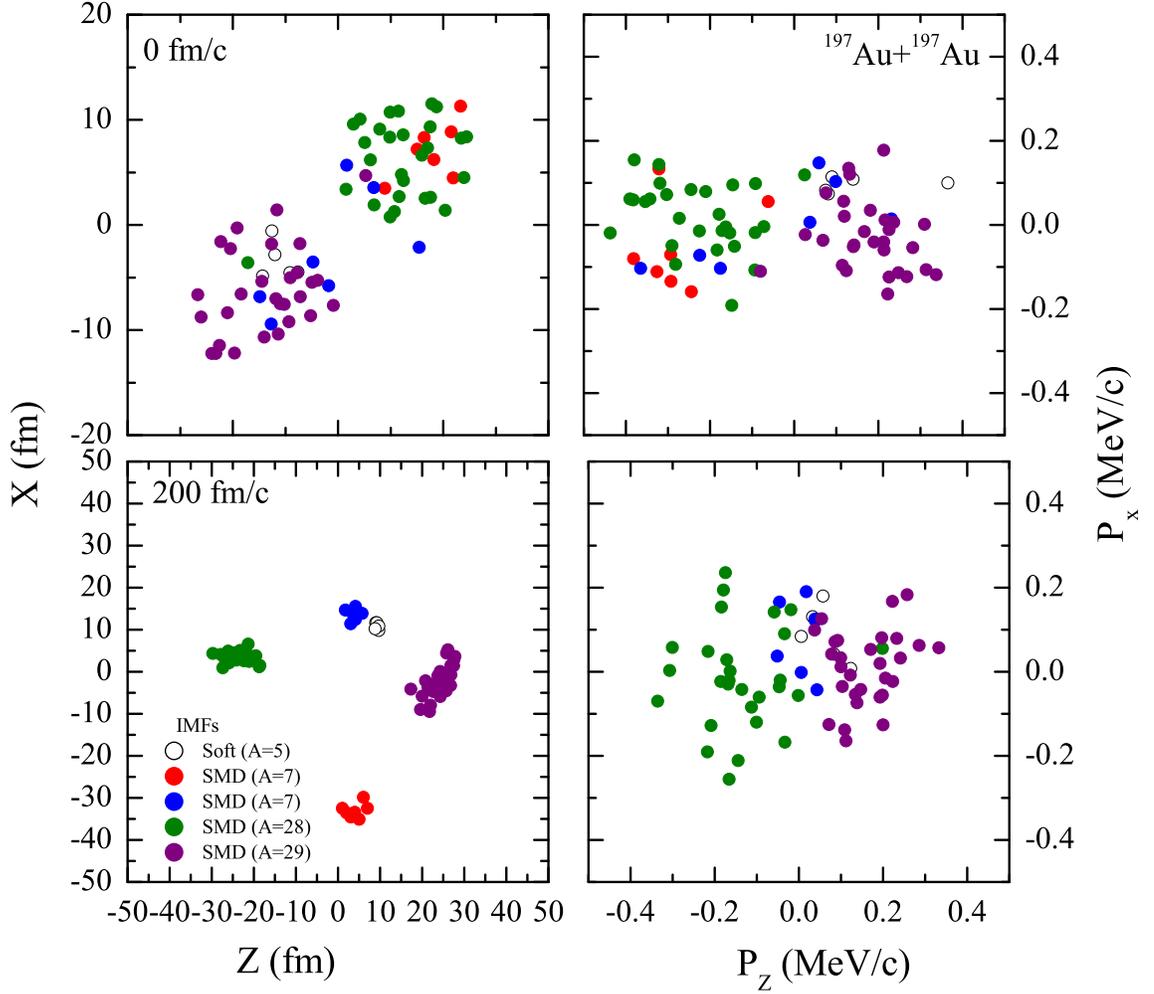}
 \vskip -2cm \caption{The phase space of the nucleons forming the fragments(IMFs) in the reaction of $^{197}$Au+$^{197}$Au
 with Soft and SMD EOS at 0 and 200 fm/c. Symbols are explained in the text.}\label{fig6}
\end{figure}

For the peripheral collision of $\hat{b}$ =0.8 (red lines), we
find that A$^{\textrm{max}}$, free nucleons, and LCPs show similar
behavior as that for central collision except that the number of
free nucleons and LCPs are now significantly reduced. This is
because of the fact that less density is achieved in peripheral
collisions and therefore, the number of IMFs is also greatly
reduced in peripheral collisions (both for soft and hard EOS) as
the static soft and hard EOS are not able to break the initial
correlations among the nucleons and hence no IMFs are emitted.
\par
In Fig. 2, we display the effect of momentum dependent
interactions on the production of A$^{\textrm{max}}$, free
nucleons, LCPs, and IMFs at $\hat{b}$ = 0.2 and 0.8. We find that
A$^{\textrm{max}}$ is nearly same for Soft and SMD (solid and
dotted line) at central collisions whereas the difference
increases at peripheral collisions. This is because in central
collisions, the nucleon-nucleon collisions are more frequent which
results in complete destruction of the initial correlations.
Therefore, an additional repulsion (due to MDI) does not alter the
results. We also see that number of free nucleons and LCPs
increases with momentum dependent interactions due to additional
destruction of the remaining correlations (at both central and
peripheral collisions). On the other hand, the role of MDI in
peripheral collisions is dominant. This is because in the
production of IMFs, the additional MDI breaks the heavy fragments
into larger number of intermediate mass fragments leading to a lot
of IMFs.
\par

\begin{figure}[!t]
\centering
 \vskip 1cm
\includegraphics[angle=0,width=16cm]{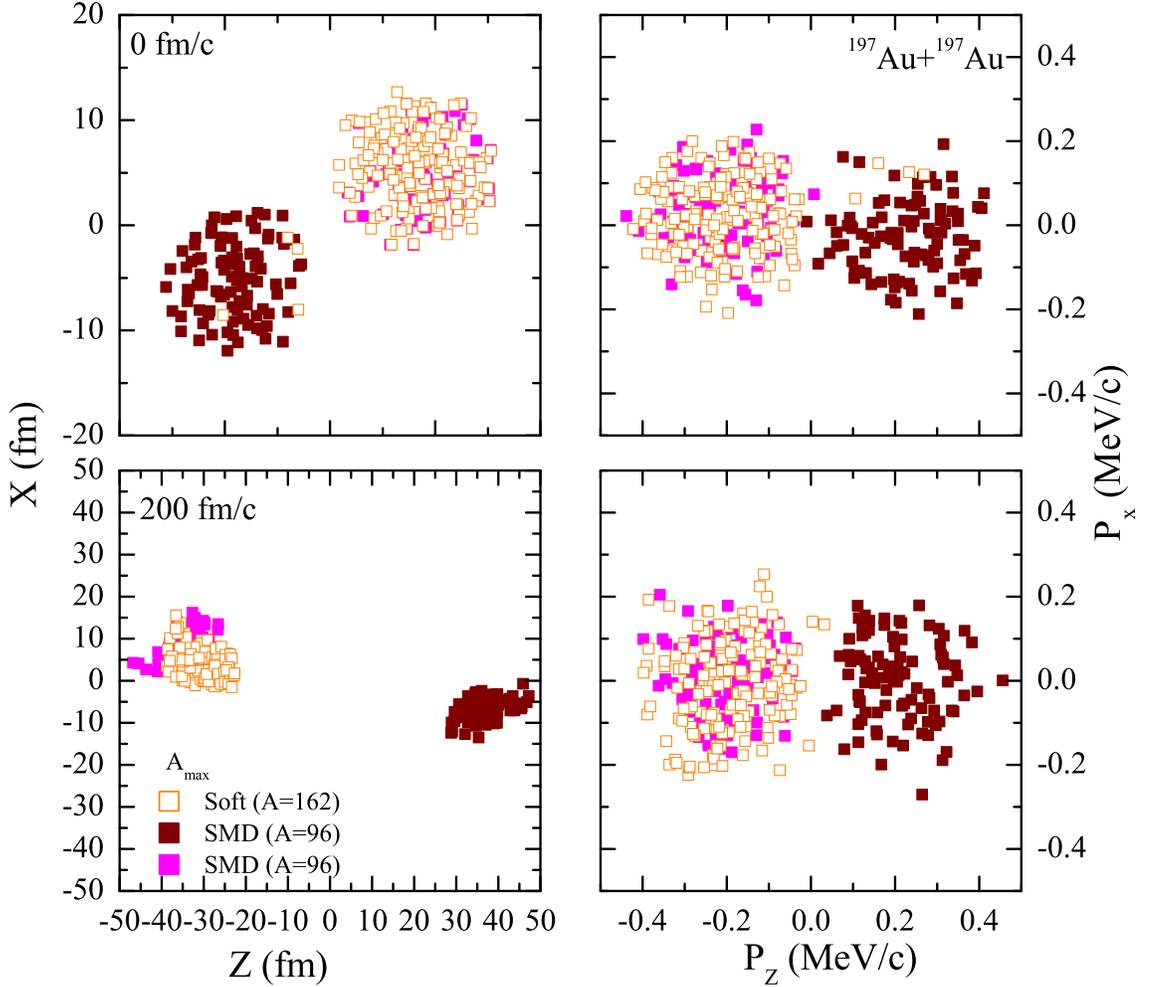}
 \vskip -2cm \caption{Same as Fig. 5, but for the A$^{\textrm{max}}$ formed with Soft and SMD EOS.}\label{fig6}
\end{figure}

In Fig. 3, we display the effect of interaction range on the
production of A$^{\textrm{max}}$, free nucleons, LCPs, and IMFs by
using two different widths of Gaussian, that is,
L$^{\textrm{norm}}$ (4.33 fm$^{2}$) and L$^{\textrm{broad}}$ (8.66
fm$^{2}$). We find that the width of Gaussian has a considerable
impact on fragmentation. As we change the Gaussian width (L) from
4.33 fm$^{2}$ to 8.66 fm$^{2}$, the multiplicity of IMFs is
greatly reduced. Owing to its largest interaction range, an
extended wave packet (L$^{\textrm{broad}}$) connects a large
number of nucleons in a fragment and as a result it generates
heavier fragments compared to what is obtained with a smaller
width. It is worth mentioning here that the width of the Gaussian
has a considerable effective on the collective flow as well as on
the pion production \cite{gaum1,hart3,klak93,kaon1}.
\par

\par
\begin{figure}[!t]
\centering \vskip 1cm
\includegraphics[angle=0,width=16cm]{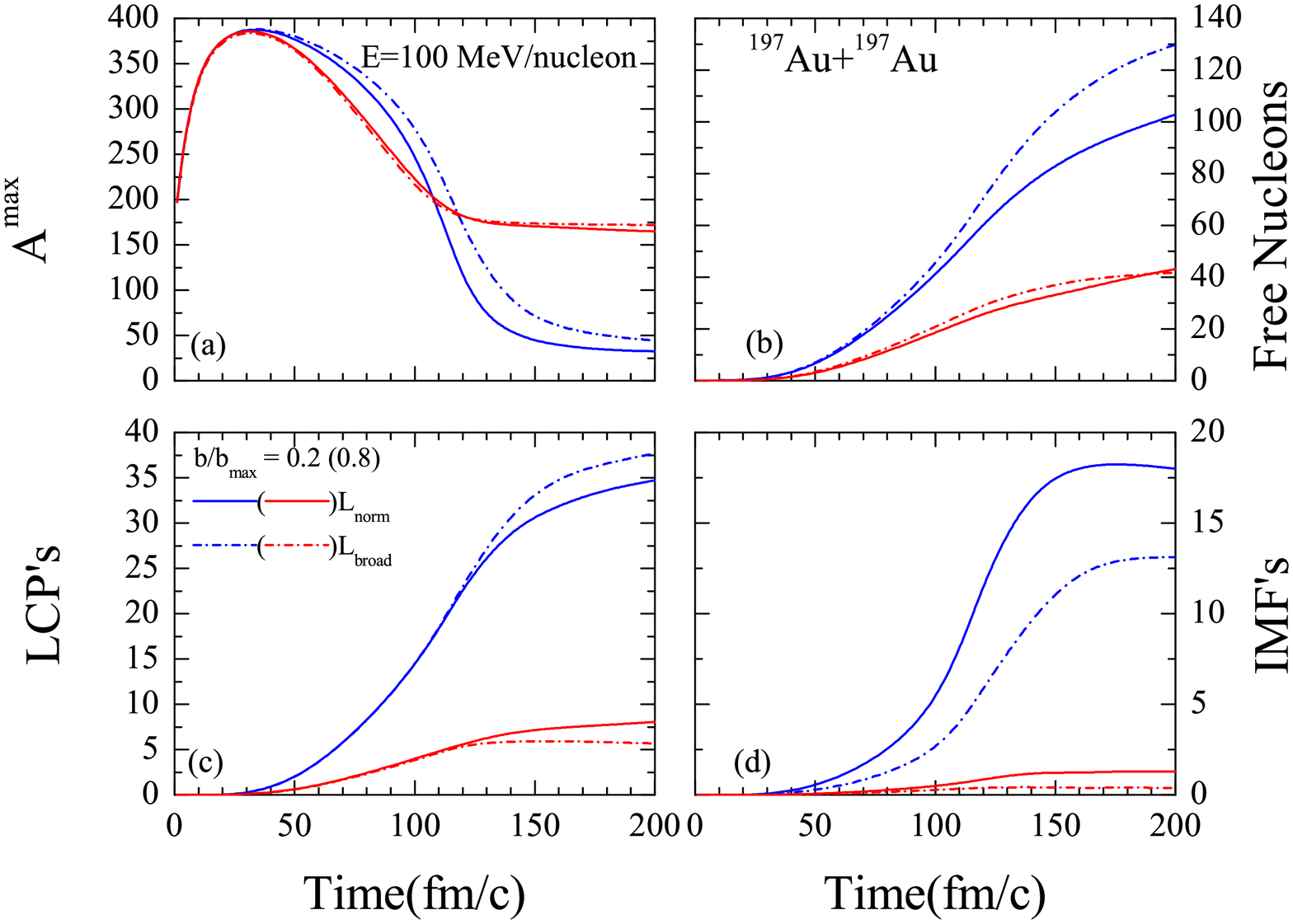}
\vskip -0cm \caption{Same as Fig. 3, but for the reaction of
$^{197}$Au+$^{197}$Au.}\label{fig3}
\end{figure}

\begin{figure}[!t]
\centering
 \vskip 1cm
\includegraphics[angle=0,width=16cm]{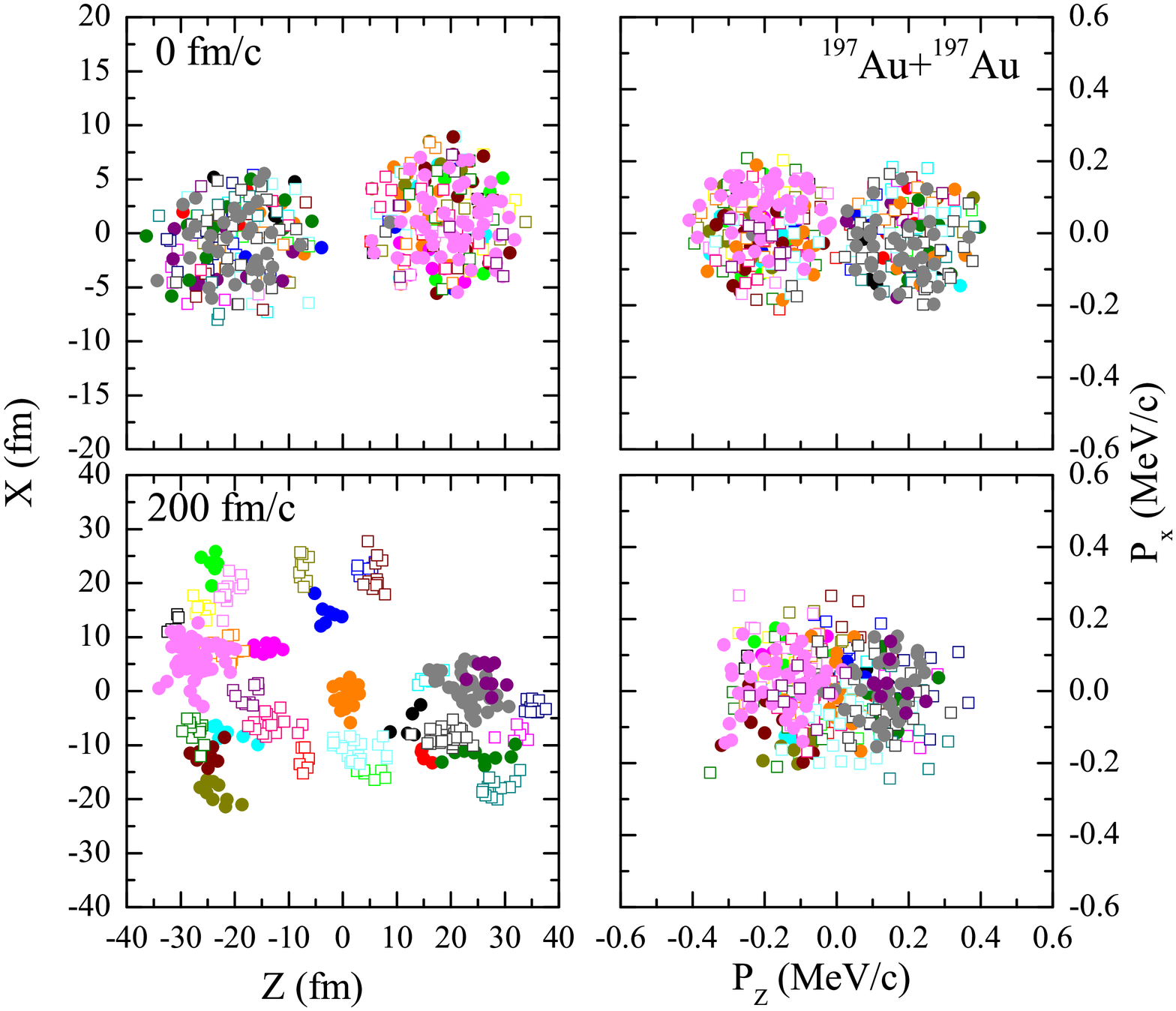}
 \vskip -2cm \caption{The phase space of the nucleons forming the fragments (IMFs) in the reaction of $^{197}$Au+$^{197}$Au
 with L$^{norm}$ and L$^{broad}$  at 0 and 200 fm/c. Solid (open) symbols are for L$^{broad}$ (L$^{norm}$). }\label{fig6}
\end{figure}

In Fig. 4 we display the effect of MDI on the reactions of
$^{197}$Au+$^{197}$Au at $\hat{b}$ = 0.2 and 0.8 for 100
MeV/nucleon. A similar behavior of all the quantities is obtained
as that for the reaction of $^{58}$Ni+$^{58}$Ni. From Fig. 4, we
see that now A$^{\textrm{max}}$ reaches a maximum value of (394)
which is the total mass of the system at the highly dense phase of
the reaction. Moreover the number of free nucleons, LCPs, and IMFs
are also increased as that in case of $^{58}$Ni+$^{58}$Ni reaction
due to increase in the system mass. From Fig. 4 we also see that
number of IMFs are more in case of SMD as that in case of soft
(static) EOS because of the destruction of initial correlations
due to the repulsive momentum dependent interactions as discussed
previously. We further investigate the details of the fragments
formed in static and MDI interactions.
\par
In Fig. 5, we display the phase space of those nucleons which form
IMFs in case of Soft and SMD EOS at initial time (0 fm/c) and the
end of reaction (200 fm/c). Left (right) panels display the
coordinate (momentum) space. Solid (open) circles represent SMD
(Soft) EOS. From the Fig., we see that nucleons forming an IMF in
case of Soft EOS belong to same region of coordinate space (see
open circles). In case of MDI also, most of the nucleons which
form the IMFs are coming from the same region (closed circles).
\par
In Fig. 6, we display the phase space of the nucleons forming the
A$^{\textrm{max}}$ in case Soft and SMD EOS. We see that for the
formation of A$^{\textrm{max}}$, the participating nucleons belong
to the same region of phase space. We also see that
A$^{\textrm{max}}$ in case of SMD is small as compared to that in
case of static one.
\par

In Fig. 7, we display the effect of interaction range on the
production of A$^{\textrm{max}}$, free nucleons, LCPs, and IMFs by
using two different widths of Gaussian, that is,
L$^{\textrm{norm}}$ (4.33 fm$^{2}$) and L$^{\textrm{broad}}$ (8.66
fm$^{2}$) for the reaction of $^{197}$Au+$^{197}$Au. We find the
similar effect of interaction range of the fragment production as
for the reaction of $^{58}$Ni+$^{58}$Ni, i.e, with broader
Gaussian, the IMF's production is reduced. To have a further
insight into the fragment structure, that is, whether the nucleons
forming a fragment when we increase the interaction range belong
to same region of phase space or not, in Fig. 8, we display the
phase space of the nucleons which are forming the IMFs both with
L$^{norm}$ and L$^{broad}$ at 0 fm/c and  200 fm/c. We find that
the nucleons which are forming the fragment belong to same region
of phase space.
\section{Summary}
We studied the effect of momentum-dependent interactions and a
broader Gaussian on multifragmentation. We also investigated the
details of the fragment structure for a broader Gaussian and
momentum-dependent interactions. We find that nucleons forming the
fragments belonged to the same region of phase space.

\section{Acknowledgement}
This work is done under the supervision of Dr. Rajeev K. Puri,
Department of Physics, Panjab University, Chandigarh, India. This
work has been supported by a grant from Centre of Scientific and
Industrial Research (CSIR), Govt. of India.

\par

\end{document}